\input harvmac.tex\overfullrule=0pt
\input mssymb.tex
\input labeldefs.tmp
\writedefs
\def\sn{\mathop{\rm sn}\nolimits}

\def\tr{\mathop{\rm tr}\nolimits}

\def\Dsl{\,\raise.15ex\hbox{/}\mkern-13mu D}
\def\Hsl{\,\raise.15ex\hbox{/}\mkern-12.5mu H}
\def\Lsl{\,\raise.15ex\hbox{/}\mkern-13mu L}
\def\Msl{\,\raise.15ex\hbox{/}\mkern-13mu M}
\def\Fsl{\,\raise.15ex\hbox{/}\mkern-13mu F}
\def\hsl{\raise.15ex\hbox{/}\kern-.57em h}
\def\omegasl{\raise.15ex\hbox{/}\kern-.57em\omega}
\def\der{\partial}

\def\d{{\rm d}}
\def\atr#1{\left<{\tr\over N}{#1}\right>}
%
\input epsf

\long\def\fig#1#2#3{%
\xdef#1{\the\figno}%
\writedef{#1\leftbracket \the\figno}%
\midinsert%
\parindent=0pt\leftskip=1cm\rightskip=1cm\baselineskip=11pt%
\centerline{\epsfbox{#3}}
\vskip 8pt\ninepoint%
{\bf Fig. \the\figno:} #2%
\endinsert%
\goodbreak%
\global\advance\figno by1%
}
%
\def\pre#1{ ({\tt #1})}
\lref\KZJ{V.A.~Kazakov and P.~Zinn-Justin, {\it Nucl.~Phys.} B 546 (1999) 647\pre{hep-th/9808043}.}
\lref\KPZ{V.~G.~Knizhnik, A.~M.~Polyakov and A.~B.~Zamolodchikov,
{\it Mod.~Phys.~Lett.~A} 3, 819--826 (1988);
F.~David,
{\it Mod.~Phys.~Lett.~A} 3, 1651--1656 (1988);
J.~Distler and H.~Kawai,
{\it Nucl.~Phys.} B 321, 509 (1989).}
\lref\ChCh{L.~Chekhov and
C.~Kristjansen, {\it Nucl. Phys.} B479 (1996), 683\pre{hep-th/9605013}.}
\lref\DFI{P.~Di~Francesco and C.~Itzykson, {\it Ann. Inst. Henri
Poincar\'e} 59, no. 2 (1993), 117.}
\lref\DK{M.R.~Douglas and V.A.~Kazakov,
{\it Phys. Lett.} B319 (1993), 219\pre{hep-th/9305047}.}
\lref\PZJ{P.~Zinn-Justin,
{\it Commun. Math. Phys.} 194 (1998), 631\pre{cond-mat/9705044}.}
\lref\KaSW{V.A.~Kazakov, M.~Staudacher and T.~Wynter,
{\it Commun. Math. Phys.} 177 (1996), 451\pre{hep-th/9502132}; 179 (1996), 235\pre{hep-th/9506174};
{\it Nucl. Phys.} B471 (1996), 309\pre{hep-th/9601069}.}
\lref\PZJb{P.~Zinn-Justin,
proceedings of the 1999 semester of the MSRI ``Random Matrices
and their Applications'', MSRI Publications Vol. 40 
(2001)\pre{math-ph/9910010}.
}
\lref\DFGK{P. Di Francesco, E. Guitter and C. Kristjansen,
{\it Nucl. Phys.} B549 (1999), 657--667\pre{cond-mat/9902082}.}
\lref\AJLV{J. Ambjorn, J. Jurkiewicz, R. Loll and G. Vernizzi,
JHEP 0109 (2001) 022\pre{hep-th/0106082}.}
\lref\IZ{C.~Itzykson and J.-B.~Zuber, {\it J. Math. Phys.} 21 (1980), 411.}
\lref\DFGZJ{P. Di Francesco, P. Ginsparg and J. Zinn-Justin, 
{\it Phys. Rep.} {\bf 254} (1995) 1--133.}
\lref\GMGW{T.~Guhr, A.~Mueller-Groeling
and H.A.~Weidenmueller,
{\it Phys. Rep.} 299 (1998), 189--425\pre{cond-mat/9707301}.}
\Title{
{\tt hep-th/0308130}
}
{{\vbox {
\centerline{The asymmetric $ABAB$ matrix model}
}}}%
\bigskip
\centerline{P. Zinn-Justin}\medskip
\centerline{\it Laboratoire de Physique Th\'eorique et Mod\`eles Statistiques}
\centerline{\it Universit\'e Paris-Sud, B\^atiment 100}
\centerline{\it F-91405 Orsay Cedex, France}
\bigskip
\noindent
In this letter, it is pointed out that the two matrix model defined by the
action
$S={1\over 2}(\tr A^2+\tr B^2)-{\alpha_A\over 4}\tr A^4-{\alpha_B\over4}\tr
B^4-{\beta\over 2} \tr(AB)^2$
can be solved in the large $N$ limit using a generalization of
the solution of Kazakov and Zinn-Justin (who considered
the symmetric case $\alpha_A=\alpha_B$). This model could have
useful applications to 3D Lorentzian gravity.

\Date{07/2003}\def\rem#1{}

\newsec{Introduction}
There are a variety of motivations to study the large $N$ limit of matrix integrals,
from statistical mechanics on random lattices and graph enumeration \DFGZJ\ to quantum chaos
and mesoscopic systems \GMGW.
In \KZJ, the so-called $ABAB$ model was introduced, as an example of model exhibiting
the critical behavior of $c=1$ conformal field theory coupled to gravity. It has the partition function
\eqn\defZsym{{\rm Z}(\alpha,\beta)=
\int\!\!\!\int \d A\, \d B\, \exp N\left[-{1\over 2}(\tr A^2+\tr B^2)
+{\alpha\over 4}(\tr A^4+\tr B^4)+{\beta\over2} \tr(AB)^2\right]
}
The special form of the interaction between the two matrices, $\tr ABAB$, which is
different from the usual two-matrix model \IZ, makes it impossible to compute this partition
function using the standard method -- direct diagonalization of $A$ and $B$. Instead a new method
was proposed, combining several techniques which had appeared in previous articles,
including character expansion \DFI, saddle point over
Young diagrams \DK\ and functional inversion relations \PZJ. 
It led to an exact expression for the ``planar'' (large $N$) free energy in terms
of elliptic functions. Subsequently, this model was used in several contexts: 2D Eulerian gravity \DFGK,
knot theory \PZJb\ and even 3D Lorentzian gravity \AJLV. This prompted the author to consider the following
natural generalization:
\eqn\defZ{{\rm Z}(\alpha_A,\alpha_B,\beta)=
\int\!\!\!\int \d A\, \d B\, \exp N\left[-{1\over 2}(\tr A^2+\tr B^2)
+{\alpha_A\over 4}\tr A^4+{\alpha_B\over4}\tr B^4+{\beta\over2} \tr(AB)^2\right]
}
in which the coefficients of the quartic terms are completely arbitrary. The Feynman rules of this model
are as follows: we form fat graphs with edges in two colors (red and green) and quartic vertices,
in such a way that colors
can only cross (each other, or themselves) at each vertex. Red (resp.\ green, mixed) vertices are assigned
a weight of $\alpha_A$ (resp.\ $\alpha_B$, $\beta$).
Alternatively, let us note that if
one sets $X={A+iB\over\sqrt{2}}$, then one obtains
\eqnn\vert
$$\eqalignno{
{\rm Z}(b,c,d,e)=\int \d X \d X^\dagger\,
\exp N\bigg[&-\tr(XX^\dagger)
+b\tr(X^2 X^{\dagger 2})
+{c\over2}\tr(XX^\dagger)^2\cr
&+{d\over4}\tr(X^4+X^{\dagger 4})
+e \tr(XX^{\dagger 3}+X^\dagger X^3)
\bigg]&\vert\cr}
$$
with $b=(\alpha_A+\alpha_B+2\beta)/4$, $c=d=(\alpha_A+\alpha_B-2\beta)/4$ and
$e=(\alpha_A-\alpha_B)/4$.\rem{anything known on this model in flat case?} It corresponds
to a perturbation of the symmetric case $\alpha_A=\alpha_B$ (which describes
an 8-vertex model) with the extra operators
$X^3 X^\dagger+{\rm h.c.}$, turning it into a 16-vertex model on random dynamical lattices. 
We shall come back
in the concluding section to possible applications of this model. For now, 
let us show how, despite the more complicated analytic structure than in the
symmetric case, one can solve it.

\newsec{Character expansion and saddle point equations}
As in \KZJ, we immediately transform the expression \defZ\ by means of a character expansion
of the term $\exp[N{\beta\over2}\tr(AB)^2]$ as a function of $AB$. We obtain:
\eqnn\charexp
$$\eqalignno{
{\rm Z}(\alpha_A,\alpha_B,\beta)=&
\sum_h (N\beta/2)^{\# h/2}
{\Delta(h^{\rm even}/2)\over \prod_i (h_i^{\rm even}/2)!}
{\Delta((h^{\rm odd}-1)/2)\over \prod_i ((h_i^{\rm odd}-1)/2)!}&\charexp\cr
&\int\!\!\!\int \d A\, \d B\, \exp N\left[-{1\over 2}(\tr A^2+\tr B^2)
+{\alpha_A\over 4}\tr A^4+{\alpha_B\over4}\tr B^4\right]
\chi_h(AB)\cr
}$$
where the sum is over a set $h=\{ h_1,\ldots,h_N\}$ of integers that satisfy
$h_1>h_2>\ldots>h_N\ge0$, and
$\# h=\sum h_i - {N(N-1)\over 2}$ is the
number of boxes of the Young diagram. $\Delta(\cdot)$ is the
Van der Monde determinant, $\chi_h$ is the $GL(N)$ character
associated to the set of shifted highest weights $h$,
and the $h^{\rm even/odd}_i$
are the even/odd $h_i$, which must be in equal numbers. 
It is now possible, using
character orthogonality relations,
to integrate over the relative angle between $A$ and $B$; this
leads to a separation into one-matrix integrals:
\eqn\charexp{
{\rm Z}(\alpha_A,\alpha_B,\beta)= \sum_h (N\beta/2)^{\# h/2} c_h
R_h(\alpha_A) R_h(\alpha_B)}
where $c_h$ is a coefficient:
$c_h={1\over\prod_i \lfloor h_i/2\rfloor ! \prod_{i,j}
(h_i^{\rm even}-h_j^{\rm odd})}$,
and $R_h(\alpha)$ is the one-matrix integral
\eqn\defR{
R_h(\alpha)=\int \d M\, \chi_h(M) \exp N\left[-{1\over 2}\tr M^2
+{\alpha\over 4}\tr M^4\right]
}
which was studied in detail in \KZJ.

We can now consider the large $N$ limit: the summation over the $h_j$ is dominated
by a saddle point which is characterized by a continuous density $\rho(h)$ 
of the rescaled variables $h_j/N$. There is generically a saturated region
$[0,h_-]$ where $\rho(h)=1$ and an unsaturated region $[h_-,h_+]$ where
$0<\rho(h)<1$. In order to write down the saddle point equation we need to differentiate
with respect to the $h_i$; we denote by $\der/\der h$ the operation $(1/N)\der/\der h_i$ in the
large $N$ limit with $h_i/N=h$. Then we have
$2(\der/\der h) \log R_h(\alpha_A)=\Lsl(h)$,
$2(\der/\der h) \log R_h(\alpha_B)=\Msl(h)$,
$(\der/\der h) \log (c_h/N^{\# h/2})=-(\Hsl(h)+\log {h\over 2})/2$,
where the slash notation means 
$\Dsl(h)={1\over2}(D(h+i0)+D(h-i0))$
for any analytic function $D(h)$ with a cut on the real axis,
$L(h)$ and $M(h)$ are unknown functions whose analytic
structure will be discussed below, and $H(h)=\lim_{N\to\infty}\sum_i {1\over h-h_i}=\int {\d h' \rho(h')\over h-h'}$.
This results in the following saddle point equation:
\eqn\spe{\Lsl(h)+\Msl(h) - \Hsl(h) = \log{h\over\beta}
\quad h\in[h_1,h_2]}

\newsec{Analytic structure and solution}
From the study of $R_h(\alpha)$ performed in \KZJ\ and that we shall not repeat here,
it follows that the function $L(h)$ defined by
$2(\der/\der h) \log R_h(\alpha_A)=\Lsl(h)$ on $[h_1,h_2]$ and by the fact that it has
the same cut as $H(h)$ on $[h_1,h_2]$, has the following analytic structure:
besides a logarithmic cut starting at $h=0$ and the cut of $H(h)$ on $[h_1,h_2]$ (i.e.\ 
discontinuity of $2\pi i \rho(h)$), it has a semi-infinite square root cut starting at
$h=h_A$, which we choose to be $[h_A,+\infty]$. 
Similary, $M(h)$ defined by replacing $\alpha_A$ with $\alpha_B$ has a cut
starting at $h=h_B$. Furthermore, they satisfy the following equations:
\eqna\anal
$$\eqalignno{
2\Lsl(h)&=H(h)+\log{h\over\alpha_A}\quad h\in[h_A,+\infty]&\anal{a}\cr
2\Msl(h)&=H(h)+\log{h\over\alpha_B}\quad h\in[h_B,+\infty]&\anal{b}\cr
}$$

Now define the function $F(h)=L(h)+M(h)-H(h)-\log h$. This is a rather complicated
function since it has 3 square root cuts leading to other sheets. 
The equations \spe\ and \anal{} define the monodromy of
$F(h)$ around the branch points $h_\pm$, $h_A$ and $h_B$. If one differentiates once $F(h)$,
the inhomogeneities in Eqs. \spe\ and \anal{} disappear (as well as the logarithmic cut)
and we find that fortunately, $F'(h)$ has only 4 sheets:
the original sheet $F'(h)$, its opposite $-F'(h)$ by going through $[h_1,h_2]$, and
$\pm G'(h)$ by going around $h_A$ or $h_B$, with $G(h)=L(h)-M(h)$. Therefore it is a meromorphic
function of $u$, where
\eqn\defu{
h-{h_A+h_B\over2}+u^2+{(h_A-h_B)^2\over u^2}=0
}
This shows that the problem is solvable.
In practice, to stay closer to the solution of \KZJ,
we introduce $z=-u^2={1\over2}(h-{h_A+h_B\over2}\pm\sqrt{(h-h_A)(h-h_B)})$, with a cut on $[h_A,h_B]$,
the $+$ sign being for the sheet of $\pm F$, the $-$ sign being for $\pm G$; also
introduce the images $z_\pm$, $z_0$, $z_A$, $z_B$ on the $F$ sheet of $h_\pm$, $0$, $h_A$, $h_B$.
Then the function $F(z)$ (which combines $F(h)$ and $G(h)$) has square root cuts on $[z_-,z_+]$ and $[0,+\infty]$ and satisfies
\eqna\speb
$$\eqalignno{
\Fsl(z)&=-\log\beta\qquad z\in[z_-,z_+]&\speb{a}\cr
\Fsl(z)&=-{1\over2} \log(\alpha_A \alpha_B)\quad z\in [0,+\infty]&\speb{b}\cr
}$$
Taking into account the extra logarithmic cut on $[z_0,z_-]$, we find that $F(z)$ is of the form:
\eqn\solF{
F(z)=-\Phi_{z_-}(z)+2\Phi_{z_0}(z)-{\log(\beta^2/\alpha_A\alpha_B)\over 2\pi i} \Phi_0(z)
}
where $\Phi_{\tilde{z}}(z)$ is the incomplete elliptic integral of the third kind, which can
be expressed explicitly as follows: define $k^2=z_+/z_-$,
the complete elliptic integrals $K$, $K'$, $E$, $E'$, and the parameterization
$\sn^2(x,k)=z_-/z$. Then 
$\Phi_{\tilde{z}}(z)=\tilde{x}(2Z_1(x)-{i\pi\over K})+\log{\Theta_1(x-\tilde{x})\over
\Theta_1(x+\tilde{x})}$, where $\Theta_a$ ($a=1,2,3,4$) is a rescaled theta function and
$Z_a$ its logarithmic derivative.

The parameters $h_\pm$, $h_A$, $h_B$ are fixed by computing the asymptotic behavior of $F(z)$
as $z\to 0$, $z\to\infty$ and the normalization of the density $\int\d h\, \rho(h)$.
This calculation is similar (though more complicated) than what is performed in \KZJ\ and we skip
the details.

First, based on the known asymptotic behavior $L(h)={1\over2} \log{-h\over\alpha_A}
-{1\over2\sqrt{-\alpha_A h}}+{1\over 2 h}+O(h^{-3/2})$ (and similarly for $M(h)$)
we obtain 3 conditions as $z\to \infty$, $0$:
\eqna\three
$$\eqalignno{
x_0&={K\over 2}+{\log(\beta^2/\alpha_A\alpha_B)\over 4\pi} K'&\three{a}\cr
\Omega_1&={1\over2\sqrt{-z_-}}(\alpha_A^{-1/2}+\alpha_B^{-1/2})&\three{b}\cr
\Omega_4&=2\sqrt{-z_+}\,{\alpha_A^{-1/2}-\alpha_B^{-1/2}\over h_A-h_B}&\three{c}\cr
}$$
where
\eqn\defom{
\Omega_a={\log(\beta^2/\alpha_A\alpha_B)\over 2K}+4Z_a(x_0)
}
and $x_0$ is defined by $\sn^2(x_0,k)=z_-/z_0$.

The normalization of the density, $\int \d h\,\rho(h)=1$, after integration by parts,
is essentially a complete elliptic integral of the second kind:
\eqn\normdens{
z_0 K'\Omega_4 - z_- E' \Omega_1 + {z_A^2\over z_0}K'\Omega_1
-{z_A^2\over z_-}{E'\over k^2} \Omega_4
=\pi}

Equations \three{}--\normdens\ provide an explicit parameterization of the quantities of the model
in terms of $k$ and ratios of coupling constants, such as $\beta^2/\alpha_A\alpha_B$
and $\alpha_A/\alpha_B$. For example, we find
\eqn\aA{
\alpha_A={1\over\pi}\left(
{E'\Omega_1-K'\Omega_4/\sn^2 x_0\over 4 \Omega_1^2} \left(1+\sqrt{\alpha_A\over\alpha_B}\right)^2
+{E' \Omega_4 - k^2 K'\Omega_1 \sn^2 x_0\over 4\Omega_4^2} \left(1-\sqrt{\alpha_A\over\alpha_B}\right)^2
\right)
}
where we recall that $x_0$ and $\Omega_a$ are known via Eqs.~\three{a} and \defom.

From the function $F(z)$ one can derive many correlation functions, including $\atr{(AB)^2}$ and
$\atr{A^n}$, but the formulae are quite cumbersome and are not reproduced here.

\newsec{Phase diagram}
Ordinary criticality, which in the context of random dynamical graphs means the emergence of very large
graphs i.e.\ the continuum limit, appears when one of the square root singularities at
$h=h_A$, $h=h_B$, $h=h_+$  degenerate into power $3/2$ singularities. Thus, there are three phases
denoted by IA, IB and II, which all correspond to pure gravity ($c=0$). 
Criticality can easily be tested numerically and the result is Fig.~\crit.
On this two-dimensional critical surface, there are critical lines 
corresponding to the transition between two of these phases and
where the corresponding branch points coalesce; and
these three lines meet at a critical point where all three branch points merge.

\fig\crit{Critical surface in the $(\alpha_A,\alpha_B,\beta)$ space.}{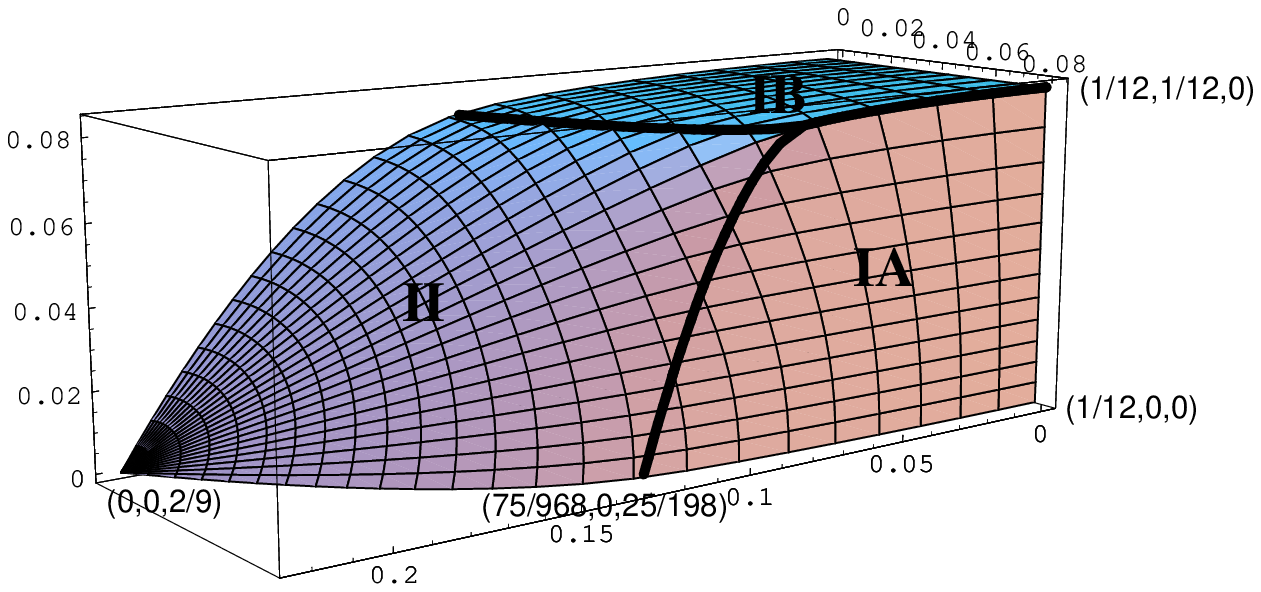}

The easiest transition to analyze is IA/IB: $h_A=h_B$ implies $\alpha_A=\alpha_B$ and we go back
to the symmetric model. According to the analysis of \KZJ,
at this transition line the model is still pure gravity.
More interesting is the IA/II transition. It is given by the condition 
$h_A=h_+$: note that contrary to the symmetric case,
$h_A=h_+$ does {\it not}\/ imply that $k\to 0$ (trigonometric limit of the elliptic functions).
However this condition still produces a degenerate analytic structure where two square root branch points
turn into a cubic branch point: this is typical of a $c=1/2$ (i.e.\ critical Ising)
theory. This is confirmed by the analysis
of the vicinity of the critical point that comes now (also, based on \ChCh\ it is reasonable
to assume that the particular point
$\alpha_B=0$ of this line is the critical point of a modified $O(1)$ model, which is known
to have $c=1/2$).

Indeed, the critical point $\alpha_A=\alpha_B=\beta={1\over 4\pi}$ was already studied in \KZJ.
It describes a $c=1$ model -- a free boson coupled to gravity.
However the vicinity, in the 3-dimensional parameter space,
displays a rather complex behavior since 3 critical lines merge there. 
Let us briefly analyze it. We define the following
deviation from criticality: $\Delta=1-{1\over4\sqrt{\pi}}(\alpha_A^{-1/2}+\alpha_B^{-1/2})$. 
The critical point
corresponds to the trigonometric limit where the elliptic nome $q$ goes to zero.
By small $q$ expansion of Eqs.~\three{}--\normdens,
we can express $\Delta$ as a function
of $q$ and of $s={\log(\beta^2/\alpha_A\alpha_B)\over 8\pi}$
and $t=(\pi\alpha_A)^{-1/2}-(\pi\alpha_B)^{-1/2}$; differentiating, we obtain
\eqn\eqDel{
{\d\over\d q} \Delta = {\log q\over 32q} {s-2q\over (s+2q)^3} (t^2 q- 64 (s+2q)^4)+\cdots
}
This shows the following scaling behaviors: $s\propto q$ and $t\propto q^{3/2}$, as well
as defines the phases in the vicinity of the critical point: $s=2q$ for phase II,
$t=\pm 8 q^{-1/2} (s+2q)^2$ for phases IA/IB. The line separating IA and II is therefore
given by $s=2q$, $t=128q^{3/2}$. Computing $\Delta$ on the critical surface allows to 
conclude that transition IA/IB is
first order, whereas transitions IA/II and IB/II are third order.\rem{this is also confirmed
by the limits $\beta=0$ and $\alpha_B$=0}

Finally, a typical correlation function $\Gamma$ follows the same pattern as $\Delta$ itself. In particular
on a generic point of 
the critical surface, ${\d\over \d q} \Gamma=0$ and therefore $\Gamma=\Delta^{3/2}+{\rm regular}$,
which confirms that the central charge $c$ is zero. However at the line separating IA and II one easily finds ${\d^2\over \d q^2}\Delta
={\d^2\over \d q^2}\Gamma=0$ and therefore $\Gamma=\Delta^{4/3}+{\rm regular}$ which according to the KPZ relation \KPZ\ 
means that $c=1/2$. 

\newsec{Concluding comments}
The asymmetric $ABAB$ model which we have just solved is enlightening in several respects. First
it is an interesting statistical model with a rich phase diagram which might
deserve further investigation. But it is also, at a technical
level, the first solution of a matrix model in which we have a non-trivial three cut structure
(the multi-cut phases of standard matrix models usually have a very simple -- e.g.\ hyperelliptic -- 
structure). It can be hoped that for many other matrix models, the seemingly complicated monodromy equations
close and allow to provide an exact solution. Finally, at the physical level we expect that the
asymmetric $ABAB$ model may be of relevance to three-dimensional Lorentzian gravity.
Indeed, it is precisely the model that was introduced in \AJLV\ (cf Eq.~(23)) to describe time slices of
three-dimensional geometries. We hope that the exact solution can shed some light on some of the
remaining unsolved issues.

\listrefs
\bye